\documentclass[twocolumn,prb,showpacs,preprintnumbers,superscriptaddress,amsmath,amssymb]{revtex4-1}
\usepackage{graphicx}
\usepackage{dcolumn}
\usepackage{color}
\usepackage{bm}
\newcommand{\bpsi}{\bar{\psi}}

\newcommand{\hh}{\hat{H}}

\newcommand{\hc}{\hat{c}}

\newcommand{\ud}{\,\mathrm{d}}

\DeclareMathOperator{\sgn}{sgn}
\DeclareMathOperator{\Order}{O}

\begin{document}

\title{Negative differential magneto-resistance in ferromagnetic wires with domain walls}

\author{N. Sedlmayr}
\email{sedlmayr@physik.uni-kl.de}
\affiliation{Department of Physics and Research Center OPTIMAS, University of Kaiserslautern,
D-67663 Kaiserslautern, Germany}
\author{J. Berakdar}
\affiliation{Department of Physics, Martin-Luther-Universit\"at Halle-Wittenberg, Heinrich-Damerow-Str. 4, D-06120 Halle, Germany}

\date{\today}

\begin{abstract}
A domain wall in a ferromagnetic one-dimensional nanowire experiences current induced motion due to its coupling with the conduction electrons. When the current is not sufficient to drive the domain wall through the wire, or it is confined to a perpendicular layer, it nonetheless experiences oscillatory motion. In turn, this oscillatory motion of the
domain wall can couple resonantly with the electrons in the system affecting the transport properties further. We investigate the effect of the coupling between these domain wall modes and the current
electrons on the transport properties of the system and show that such a system demonstrates negative differential magnetoresistance due to the resonant coupling with the low-lying modes of the domain wall motion.
\end{abstract}


\maketitle

\section{Introduction}

When a current is passed through a domain wall (DW) in a ferromagnetic wire then the spin orientation of the current electrons is changed as they pass through the DW, both due to adiabatic changes and scattering from the noncollinear magnetization. Conservation of angular momentum tells us that this must be compensated. Some of the angular momentum is transferred to the lattice and some is compensated by a change in the angular momentum of the magnetization comprising the DW. This change in angular momentum effectively sets the wall in motion, which can be described using, for instance, the Landau-Lifschitz-Gilbert (LLG) equation.\cite{ll9,gilbert04,zhang04,slonczewski} The details of the carrier-DW coupling, and the subsequent magnetization dynamics of the DW set into motion by the applied current have recently been the subject of a large volume of work,\cite{marrows05} which may also have possible applications to memory device technology.\cite{parkin08} The interaction of the current electrons with the domain wall can even induce interaction effects between the DWs.\cite{sedlmayr09,Sedlmayr2011a,Sedlmayr2011c,PhysRevLett.96.047208}

In this paper we wish to address effects which occur when the domain wall is pinned. If the current is not large enough to cause lateral motion of the DW through the wire it is nevertheless possible for the DW to experience oscillatory motion.\cite{winter61,thiele73,rebei06,sedlmayr09,kiselev} Such oscillations of the DW have been experimentally verified\cite{saitoh04} and can be thought of as the zeroth mode excitations of the DW. In this paper we describe the effect that the coupling between such DW modes and the conduction electrons has on the current through the wire. The resonant coupling between the conduction band and the mode of the domain wall's motion leads to a negative differential magnetoresistance (NDMR). Negative differential resistance (NDR) is a well known  phenomenon \cite{NDR_book,chang} that has been
studied extensively, e.g.  for semiconductor-based resonant-tunneling diodes, for organic semiconductor spin valve systems,\cite{wei06,wei07} and for double quantum dot set-ups.\cite{trocha09} NDR has found diverse applications,  e.g. for microwave signal amplifications,\cite{microwave} in feedback oscillators,\cite{oscillator} in
frequency mixers,\cite{mixer} and others.
To our knowledge, however, a DW negative differential \emph{magneto}-resistance has not been investigated yet.
In view of the NDR applications mentioned above and  the fact that  domain wall logics is well-established by now,\cite{dwlogic} it is timely to consider NDMR in noncollinear magnetic structures.

The model we will consider is also appropriate for considering a three part set up consisting of two ferromagnetic wires of opposite magnetic orientation with a strip of ferromagnetic material between them. The middle layer will have a magnetization perpendicular to the two wires and as a current is passed through a spin torque also acts on this thin layer causing its magnetic orientation to move.\cite{slonczewski} Such structures are also experimentally promising from our point of view.

In addition to the coupling between the DW mode and the electrons there are  contributions to the resistance from scattering from the domain wall.\cite{wickles09,levy97,dugaev02,tatara97} These contributions to the current are included exactly within our model and are naturally the cause of the DW's motion. The scattering of the electrons from the DW leads to a charge and spin build up in its vicinity and Friedel oscillations in the spin-dependent density of the carrier electrons. This build up of spin enhances the coupling between the DW motion and the electrons.

Our method is to first solve the effect of the DW and its motion on the charge build up in the system. The motion of the domain wall is modeled by the DW oscillatory mode  coupled to the electrons in the vicinity of the DW.\cite{winter61,thiele73,wei06,wei07} This oscillatory mode is described by a free energy for the classical magnetization.

\section{Model}

We start from a model for the classical inhomogeneous magnetization that we describe by
a time dependent unit vector field $\vec{n}(\vec{r},t)$ since we assume the longitudinal dynamics
are energetically forbidden.
We include the effects of the anisotropy and exchange and the coupling of $\vec{n}(\vec{r},t)$
 with the conduction electrons with a Kondo-type coupling strength $J$. $\vec{T}$ is an applied field modeling the torque caused by the conduction electrons and $\vec{S}$ is the spin density of the conduction electrons.
  The magnetization  free energy then reads \cite{ll9}
\begin{eqnarray}
\mathcal{F}&=&\int\ud^3\vec{r}\bigg[f'_0\vec{n}-\alpha_{ik}\frac{\partial^2\vec{n}}{\partial r_i\partial r_k}-\vec{T}-J\vec{S}
-Kn_\beta\hat{\bm{\beta}}\bigg].\vec{n}.\nonumber\\&&
\end{eqnarray}
$f_0$ is the homogeneous part of the exchange energy. $\alpha_{ik}$ is a tensor describing the inhomogeneous part of the exchange energy and is of order $\sim T_c/a$ where $T_c$ is the Curie temperature and $a$ is the lattice spacing.
$K$ gives the anisotropy, taken in the $\hat{\beta}$ direction.

For the conduction electrons we have, in addition to the coupling to this bulk magnetization, the Hamiltonian (we set  $\hbar=1$ throughout)
\begin{eqnarray}
\hh_{e}&=&\int \ud^3 \vec{r}\sum_{\sigma}\hc^\dagger_{\sigma}(\vec{r})[\hat{\varepsilon}-\mu]\hc_{\sigma}(\vec{r})
\end{eqnarray}
where $\hat{\varepsilon}$ is the single particle dispersion
measured with respect to the Fermi level $\mu$ and $\hc_{\sigma}(\vec{r})$ are the carrier field operators with a spin index $ \sigma$.
 The total system is then described by $\mathcal{F}+\hh_e$.

Upon perturbation, the magnetization  $\vec{n}(\vec{r},t)$  undergoes some periodic motion in the vicinity of the pinned DW.
 The coupling between the motion of the DW and the conduction electrons can be  treated as a bosonic mode of the system.
Thus, integrating over the spatial directions perpendicular to the wire, we can write down the following one dimensional Hamiltonian:
\begin{eqnarray}
\hh&=&\hh_0+\int\ud z\sum_{\sigma \sigma'i}\hc^\dagger_{\sigma}(z)g_{\sigma\sigma'}(z)\big(\phi_i+\phi_i^\dagger\big)\hc_{\sigma'}(z)\nonumber\\&&\qquad+\sum_i\varepsilon_{ir}\phi_i^\dagger\phi_i,
\end{eqnarray}
where $\varepsilon_{ir}$ are the energies of the DW's resonant motion, $g_{\sigma\sigma'}(z)=g_{\sigma\sigma'}\delta(z)=g\sigma^x_{\sigma\sigma'}\delta(z)$ is the coupling between the DW mode and the electrons, and $\phi$ models the DW mode. $\hh_0$ is the ``uncoupled'' part of the Hamiltonian. It is important, for the possibility of observing a negative differential resistance effect, that the energy of the mode of motion of the DW is of the order of the carrier electrons' energy. This requires either rather narrow domain walls or a suitable layer of magnetic noncollinearity. Here we consider a system with a relatively low carrier density such that their Fermi wavelength $\lambda_F\gtrsim L$, where $L$ is the DW width. Therefore it is natural to model the system with sharp domain walls. This has the further advantage of analytic tractability. The carrier scattering from a DW has been addressed previously.\cite{araujo06} From the transmission and the reflection coefficients we find the charge build up in the system due to the DW in the presence of an applied bias. In this case we deal with a standard $s-d$ Hamiltonian
\begin{eqnarray}\label{h2}
\hh_0&=&\int \ud z\sum_{\sigma}\hc^\dagger_{\sigma}(z)[\hat{\varepsilon}-\mu]\hc_{\sigma}(z)\\&&-J\int\ud z\nonumber \sum_{\sigma\sigma'}\hc^\dagger_{\sigma}(z)[n_z(z)\sigma^z_{\sigma\sigma'} +\lambda\delta(z)\sigma^x_{\sigma\sigma'}]\hc_{\sigma'}(z).
\end{eqnarray}
Where
\begin{eqnarray}
\lambda=\int_{-\infty}^{\infty}\ud zJn_x(z)\approx J\pi L\textrm{ and}\\
\lim_{L\to 0}Jn_z(z)\to-J\sgn(z).
\end{eqnarray}
Without loss of generality we have orientated our domain wall in the x direction. We have implicitly assumed a separation of time scales which allows us to treat the domain wall as adiabatic on the time scale relevant for the dynamics of the conduction electrons.

\section{Calculation}

The electronic Green's function is
\begin{eqnarray}
i\mathbf{G}(z,z';t,t')=\frac{1}{\mathcal{Z}}\int D\psi D\bpsi D\phi D\bar{\phi} {\bm\psi}_{z}(t){\bm \bpsi}_{z'}(t')
e^{iS}
\end{eqnarray}
with the action, $S=S_{\psi}+S_{\phi}+S_{\phi\psi}$, given by
\begin{eqnarray}
S_{\psi}&=&\int_c \ud t\ud z\sum_{\sigma\sigma'}\bpsi_{\sigma}(z,t)\left[i\partial_t\delta_{\sigma\sigma'}-\hat{H}_0\right]\psi_{\sigma'}(z,t)
\nonumber\\
S_{\phi\psi}&=&-\int_c \ud t\ud z\sum_{\sigma\sigma'}\bpsi_{\sigma}(z,t)\left[g_{\sigma\sigma'}(z)\left(\phi+\bar{\phi}\right)\right] \psi_{\sigma'}(z,t)
\nonumber\\
S_{\phi}&=&\int_c\ud t \bar{\phi}\underbrace{[i\partial_t-\varepsilon_{ir}]}_{=D^{-1}_i(t)}\phi.
\end{eqnarray}
For convenience $c$ is the full interaction time contour.\cite{Sedlmayr2006} One can fully integrate out the bosonic degree of freedom, and then expand perturbatively in the coupling $g$. Switching to the Keldysh representation\cite{rammer86,kamenev05} and introducing the following retarded/advanced bosonic Green's functions, $P^{R/A}(\omega)=D^{R/A}(\omega)+D^{R/A}(-\omega)$,
gives the first order correction in perturbation theory:
\begin{eqnarray}\label{apertg}
G_{\substack{in\\\sigma\sigma'}}(z,z';\varepsilon)\approx  G^0_{\substack{in\\\sigma\sigma'}}(z,z';\varepsilon)\qquad\qquad\qquad\qquad\qquad\quad\\
+i\int \frac{\ud \omega}{2\pi}\ud z_1\ud z_2g_{\sigma_1\sigma_2}(z_1)g_{\sigma_3\sigma_4}(z_2)\gamma^s_{jk}\gamma^p_{lm}\nonumber\\
\bigg[P^{sp}(\omega) G^0_{\substack{ij\\\sigma\sigma_1}}(z,z_1;\varepsilon) G^0_{\substack{kl\\\sigma_2\sigma_3}}(z_2,z_3;\varepsilon-\omega)
G^0_{\substack{mn\\\sigma_4\sigma'}}(z_4,z';\varepsilon)
\nonumber\\
-P^{sp}(0) G^0_{\substack{im\\\sigma\sigma_3}}(z,z_2;\varepsilon) G^0_{\substack{ln\\\sigma_4\sigma'}}(z_2,z';\varepsilon) G^0_{\substack{kj\\\sigma_1\sigma_2}}(z_1,z_1;\omega)\bigg].\nonumber
\end{eqnarray}
We use Latin indices for the Keldysh matrix indices. Summation over repeated indices is implied. The emission/absorption tensors are $\gamma^1_{ik}=\delta_{ik}$ and $\gamma^2_{ik}=\sigma^x_{ik}$.

\begin{figure}
\includegraphics*[width=0.48\textwidth]{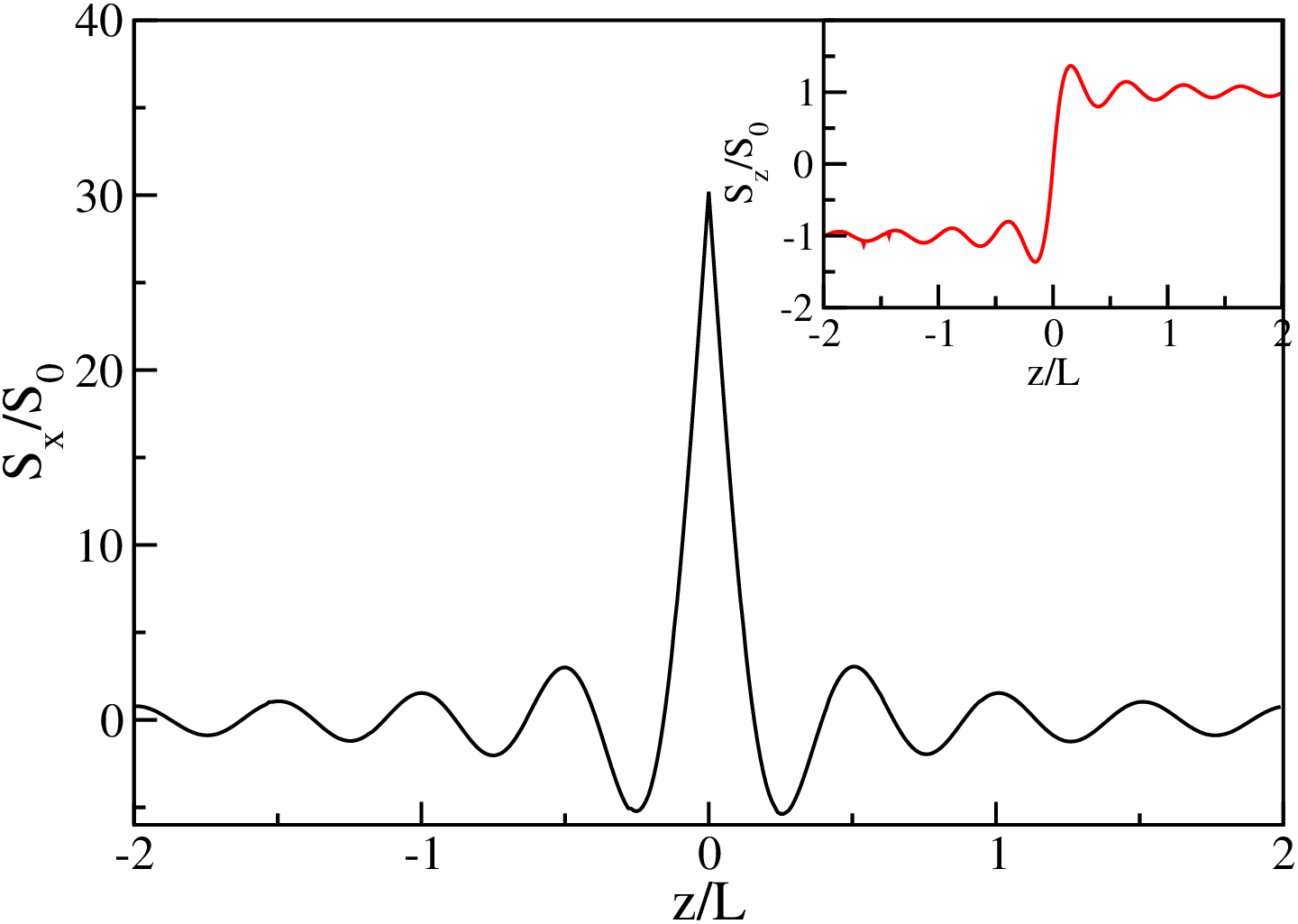}
\caption[Spin Density]{(Color online) The $x$- and the $z$ component (inset) of the spin density $\vec{S}(z)$, around the domain wall (of width $L$) caused by scattering of electrons from the wall.}\label{spindensity}
\end{figure}

The first step is to calculate the spin density of the conduction electrons which couples to the magnetization, determined by Eq.~\eqref{h2}. It is helpful to first decompose  into the scattering states $\psi_{n}(z,\sigma)$ labeled by $n_\pm=\{\pm k_\varepsilon,\alpha\}$.\cite{dugaev03} ($\alpha$ is the spin  and $\pm k_\varepsilon$ the momentum of the incoming electron.) Then the spin density becomes $\vec{S}(z)=\vec{S}_L(z)+\vec{S}_R(z)$ where
\begin{eqnarray}
\vec{S}_{L,R}(z)&=&\sum_{\substack{\sigma\sigma'\\\alpha}}\int^{\varepsilon_F\pm\frac{eV}{2}}\ud\varepsilon\nu_\alpha(\varepsilon)\psi^*_{n_\pm}(z,\sigma)
\psi_{n_\pm}(z,\sigma')\vec{\sigma}_{\sigma\sigma'}.\nonumber\\&&
\end{eqnarray}
The $L$ and $R$ indices naturally refer to electrons incident from the left and right, which are held at a potential drop of $eV$ symmetrically across the Fermi energy $\varepsilon_F$. $\nu_\alpha(\varepsilon)$ is the density of states of the carriers. In the region around the DW the transpose component of the spin density is enhanced, see Fig.~\ref{spindensity}. Density fluctuations are also clearly visible.

The uncoupled Green's function in the scattering basis becomes
\begin{eqnarray}\label{gscat}
G^0_{\sigma\sigma'}(z,z';\varepsilon)&=&\sum_{n} \frac{\psi_n(z,\sigma)\psi^\dagger_n(z',\sigma')}{\varepsilon-\varepsilon_n+\mu_c+i\delta\sgn(\varepsilon)},
\end{eqnarray}
using the shorthand
\begin{eqnarray}
\sum_n\ldots&=&\sum_{c=\{L,R\}}\sum_{\alpha}\int \ud \varepsilon_n\nu_{\alpha}^{c}(\varepsilon_n)\ldots
\end{eqnarray}
with the dispersion of the incoming electrons' $\varepsilon_n$. The chemical potential $\mu_c$ will also depend on the scattering state as the electrons incoming from the left/right have a different chemical potential: $\varepsilon_F\pm eV/2$, respectively.

Ultimately we are interested in the current through the system:
\begin{eqnarray}
\hat{j}_\sigma(z)=\frac{e}{m}\Im\hat{\psi}^\dagger_\sigma(z)\frac{\partial}{\partial z}\psi_\sigma(z),
\end{eqnarray}
for spin $\sigma$ electrons. We are interested in the steady state case where
 there is no charge build up in the wire. The current is then
\begin{eqnarray}
I=\sum_\sigma\langle\hat{j}_\sigma\rangle=-\sum_\sigma\frac{e}{m}\Re\partial_z G^<_{\sigma\sigma}(z,z';t,t^+)|_{z'=z}.
\end{eqnarray}
We can substitute in our perturbative expression Eq.~\eqref{apertg}, and Eq.~\eqref{gscat} and work in the limit of zero temperature.

\section{Results}

\begin{figure}
\includegraphics*[width=0.48\textwidth]{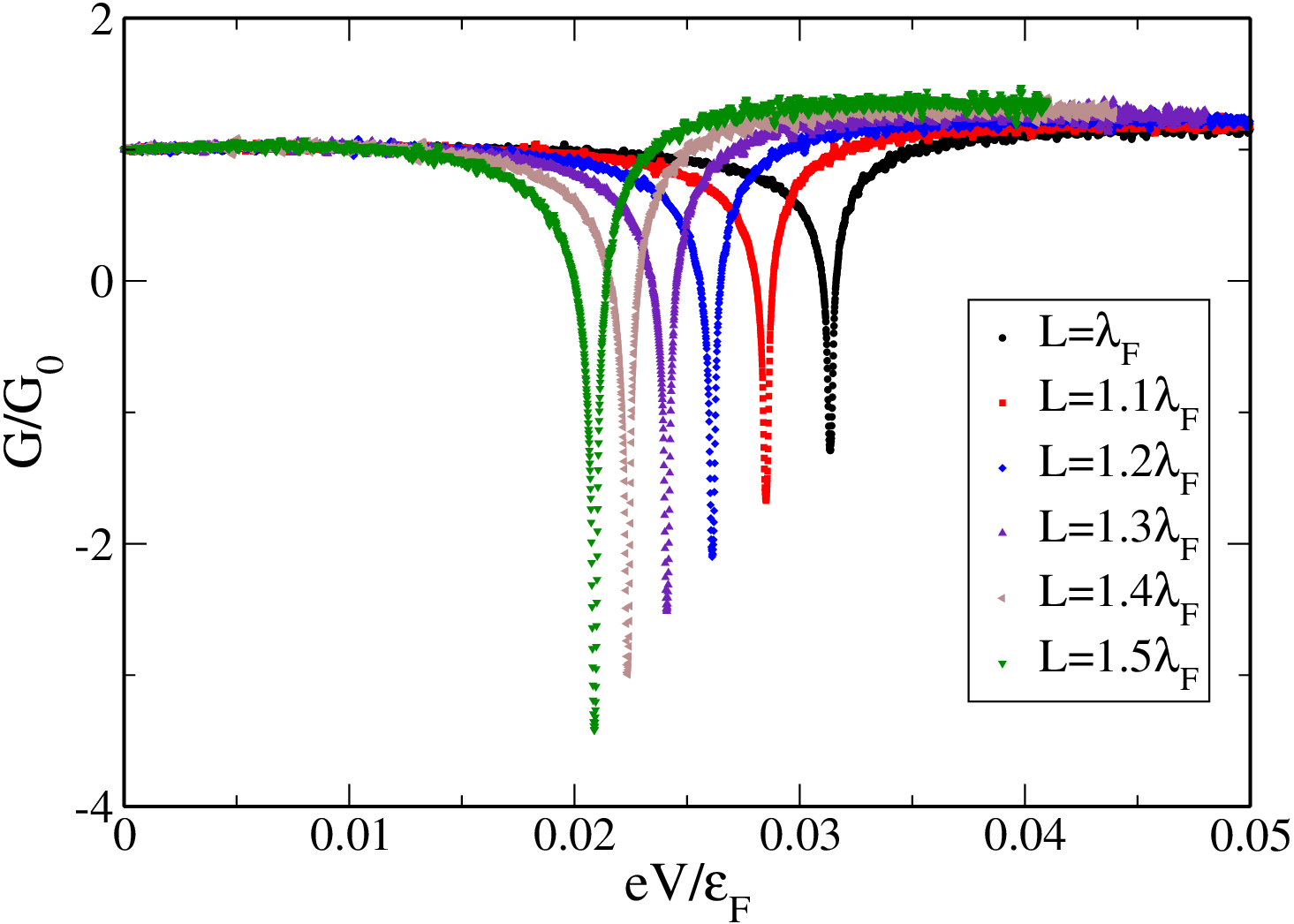}
\caption[Current]{(Color online) The differential conductance $G(V)$ normalized to $G_0\equiv G(V,g=0)$, as a function of the bias voltage for
different widths $L$ of DW in units of the carrier wave length $\lambda_F$.
The negative differential conductance occurs around $\varepsilon_r$.}\label{fig_diff_cond}
\end{figure}

\begin{figure}
\includegraphics*[width=0.48\textwidth]{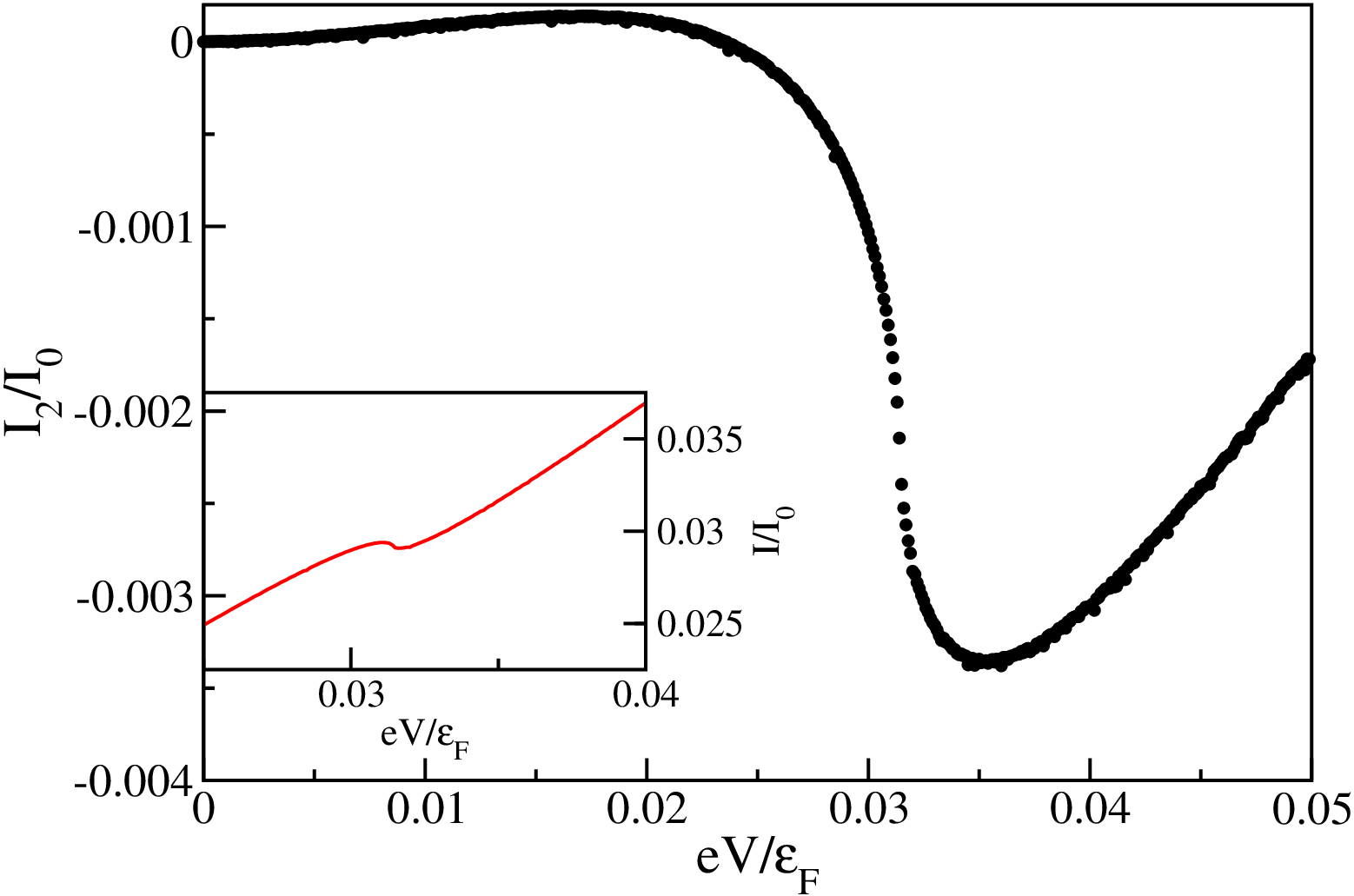}
\caption[Current]{(Color online) The current $I(V)$  as a function of the bias voltage. $I$ is measured in units of $I_0\equiv I(eV=\varepsilon_F,g=0)$. The inset shows the total current, where the feature is dimly visible. In the main figure we plot the second order correction to $I(V)$: $\Order(g^2)$. Here the drop in current due to the resonant mode coupling is visible. Note however that of course the differential conductance is the experimentally relevant quantity.}\label{fig_current}
\end{figure}

The results are plotted  in Figs.~\ref{fig_diff_cond} and \ref{fig_current}, the negative differential conductance is marked by a dip in the current around the lowest resonance $\varepsilon_{0r}=\varepsilon_r$. We consider magnetic semiconductor wires \cite{ruster03,sugawara08,matsukara} where the DW width can be closer to the Fermi wavelength, the width of the wall can be atomically sharp  when constrictions are present \cite{bruno99,pietzsch00,ebels00}. Hence we take $\lambda_F\approx L$, $J=0.005\varepsilon_F$. The mode energy is $\varepsilon_{r}=2\pi J\sigma_{cs}/La$, where $\sigma_{cs}$ is the cross section and $a$ is the lattice spacing. To keep this energy within experimentally reasonable sizes, we consider quasi-one-dimensional wires where $\sigma_{cs}\sim L^2$. This gives us $\varepsilon_{r}\approx 0.03\varepsilon_F$\cite{thiele73}, and for the unrenormalized coupling we take $g_0=10JL$
At low temperatures in such systems electron-correlation effects tend to renormalize the coupling strengths.\cite{araujo06,sedlmayr2011b} Hence we take the effective coupling as $g_{eff}\sim g_0\xi^{1-\gamma}$, with $\xi\sim\beta J=5$ and $\gamma=0.8$. The scattering strengths from the DW will be similarly renormalized.

Of experimental relevance is the differential conductance which can be defined as $G(V)=\ud I/\ud V$. This is plotted in Fig.~\ref{fig_diff_cond} for different DW widths, and the negative differential conductance occurs around the resonant energy level $\varepsilon_r$. Higher modes of the DW's motion would couple similarly at different energies for larger applied bias.

We note that  the model developed in this work  is also  relevant to interacting quantum dot systems involving regions with
 a noncollinear magnetization.\cite{Sedlmayr2006,Sedlmayr2008} This is the subject of ongoing work.

With regard to an experimental realization we note the following:   NDMR is sizable for DWs in magnetic-semiconductor-based nanowires such as those realized in  Refs.~\onlinecite{ruster03,sugawara08,matsukara}.
We developed the above theory  however for electrons as carriers whereas  in III-V
compounds the carriers  are
 holes. As follows from the above treatment,
 the underlying mechanisms for the emergence of NDMR
  are however the DW scattering and inferences of the carriers coupled to the
  modes of the DW.  These elements will also be present for
 carrier holes  despite their more complicated electronic dispersion, and hence we expect
  NDMR to also be present in this case.

\section{Summary}

In summary, we have calculated the current through a ferromagnetic wire with a domain wall present, or equivalently a thin ferromagnetic strip between ferromagnetic wires of opposing magnetization. Particular attention has been paid to the effects of the lowest modes of the DWs motion caused by the presence of this current. In order to solve for the density build up around the domain wall subject to an applied potential we used a soluble model for a sharp DW. The negative differential magnetoresistance expected from the excitation of these modes by the current electrons is clearly visible in the differential conductance depicted in Fig.~\ref{fig_diff_cond}. This qualitatively new effect
 offers  new opportunities for applications in spintronics and domain wall logics \cite{dwlogic} similar to those that utilize NDR in charge-based electronics.

\section*{Acknowledgments}
We thank V. Dugaev for many valuable discussions. This work is  supported by the DFG contract BE
2161/5-1, and SFB 762, and by the Graduate School of MAINZ (MATCOR).


\end{document}